\documentclass[final,5p,times,twocolumn]{elsarticle}
\usepackage{graphics}

\usepackage{lineno}

\def\pbdd{$^{210}$Pb }

\def\podd{$^{210}$Po }
\def\poddn{$^{210}$Po}
\def\udt{$^{238}$U }

\def\thdt{$^{232}$Th }
\def\thdtn{$^{232}$Th}
\def\tld{$^{208}$Tl }
\def\tldn{$^{208}$Tl}
\def\bidq{$^{214}$Bi }

\def\kqn{$^{40}$K}

\def\amnu{$\vert\langle m_{\nu} \rangle\vert$~}

\def\BBz{$\beta\beta(0\nu)$~}
\def\BBzn{$\beta\beta(0\nu)$}

\def\BBd{$\beta\beta(2\nu)$~}
\def\BBdn{$\beta\beta(2\nu)$}
\def\BB{$\beta\beta$~}

\def\ca{$\sim$}

\def\teod{TeO$_2$~}
\def\teodn{TeO$_2$}
\def\be{\begin{equation}}
\def\ee{\end{equation}}

\def\less{$<$}

\journal{Atroparticle Physics}
\begin{document}

\begin{frontmatter}

\title{Characterization of ZnSe scintillating bolometers for Double Beta Decay}

\author[INFN]{C.~Arnaboldi} 
\author[UNIMIB,INFN]{S.~Capelli}
\author[INFN]{O. Cremonesi}
\author[UNIMIB,INFN]{L. Gironi}
\author[UNIMIB,INFN]{M. Pavan \corref{cor1}}\ead{maura.pavan@mib.infn.it}
\author[INFN]{G. Pessina}
\author[INFN]{S. Pirro}

\address[INFN]{INFN - Sezione di Milano Bicocca I 20126 Milano - Italy}
\address[UNIMIB]{Dipartimento di Fisica - Universit\`{a} di Milano Bicocca I 20126 Milano - Italy}

\date{\today}
\begin{abstract}
ZnSe scintillating bolometers are good candidates for future Double Beta Decay searches, because of the $^{82}$Se high Q-value and thanks to the possibility of alpha background rejection on the basis of the scintillation signal. In this paper we report the characteristics and the anomalies observed in an extensive study of these devices. Among them, an unexpected high emission from alpha particles, accompanied with an unusual pattern of the light vs. heat scatter plot. The perspectives for the application of this kind of detectors to search for the Neutrinoless Double Beta Decay of $^{82}$Se are presented.
\end{abstract}

\begin{keyword}

Double Beta Decay \sep Bolometers \sep ZnSe \sep Quenching Factor 
\PACS 23.40B  \sep 07.57K \sep 29.40M

\end{keyword}

\end{frontmatter}

\section{Introduction} \label{sec:intro}
Neutrinoless Double Beta Decay (\BBzn) is, at present, a unique method to investigate basic neutrino properties. The observation of this transition would give insight on the mass pattern of the three neutrino eigenstates and would asses the Majorana character of neutrinos~\cite{reviewDBD}.
The purpose of new generation \BBz experiments \cite{gerda,CUORE,majorana,exo,supernemo} is to reach a sensitivity on \amnu of the order of~50 meV, approaching - but probably not completely covering - the so called \emph{inverse hierarchy region}. The realization of an experiment with a reasonable discovery potential down to the smallest \amnu values allowed by the inverted hierarchy (\ca~10 meV), is today an incredible challenge in which detector technology plays a key role.

In this paper we discuss the technical results obtained with ZnSe cryogenic particle detectors and the perspectives of their application to the study of \BBz decay of the $^{82}$Se isotope.

\section{Scintillating bolometers}\label{sec:bolux}

Cryogenic particle detectors (also named bolometers) have been proposed since many years for the study of rare events such as \BBz and Dark Matter \cite{bolometers}. The excellent detector performance and Physics results obtained by the CUORICINO experiment \cite{cuoricino} prove how this technique is competitive in the case of \BBz searches.

 Unlike other solid state devices, bolometers are not based on ionization but on phonon detection. A bolometer is made of three elements: an \emph{absorber} where the energy of the particles is deposited and converted into phonons, a \emph{phonon sensor} that transduces the phonon signal into an electrical pulse and a \emph{thermal link} to a low temperature (below 100 mK) heat sink that maintains the detector at the optimal temperature.
 
  The working principle of these devices can be summarized as follow. By interacting in the absorber, a particle produces ionization and excitation, which starts triggering a series of processes that lead to the conversion of almost all the deposited energy into phonons (i.e. lattice elementary excitations). Along this path it is possible that a fraction of the energy is trapped in long-living excitated states or is lost through the emission of photons that escape the absorber. However, a proper choice of the absorber (usually a single dielectric crystal, with a high Debye temperature and without scintillation properties) allows to reduce to a minimum the importance of these channels and therefore ensures a high efficiency in the conversion of the deposited energy into phonons. 
  The phonon signal is detected through a sensor that modifies its electrical properties when collided with a phonon flux. 
The dimensions and characteristics of the absorber and of the phonon sensor define the specific performance of the bolometer as far as energy resolution and time-response is concerned. 
 
 In the bolometers described in this work the sensor is a semiconductor thermistor glued on the crystal, while the absorber is a large size dielectric crystal (whose heat capacity is expected to dominate over the thermistor one). In this configuration the detector response is extremely slow: the typical rise time is of the order of few ms.
 Also CUORICINO detectors are calorimeters of this kind. They are crystals of \teodn, with a mass of 790~g. Their energy resolution at the \BBz transition is 7 keV FWHM, quite similar to the performance of a Ge diode. However, when compared with the more competitive devices used today for \BBz decay investigations, it is evident that the main disadvantages of bolometers is the lack of an active background rejection (e.g. particle identification as used in NEMO \cite{NEMOdetector} or single-site vs. multi-site events distinction as in GERDA and Majorana \cite{gerda,majorana}). The use of scintillating crystals, as absorbers, can overcome this limitation \cite{nsvecchioarticoloCaF2,CRESST,ROSEBUD,Pirr06}, providing a powerful tool to improve the experimental \BBz sensitivities.

\begin{figure}
\begin{center}
\includegraphics[ width=1\linewidth]{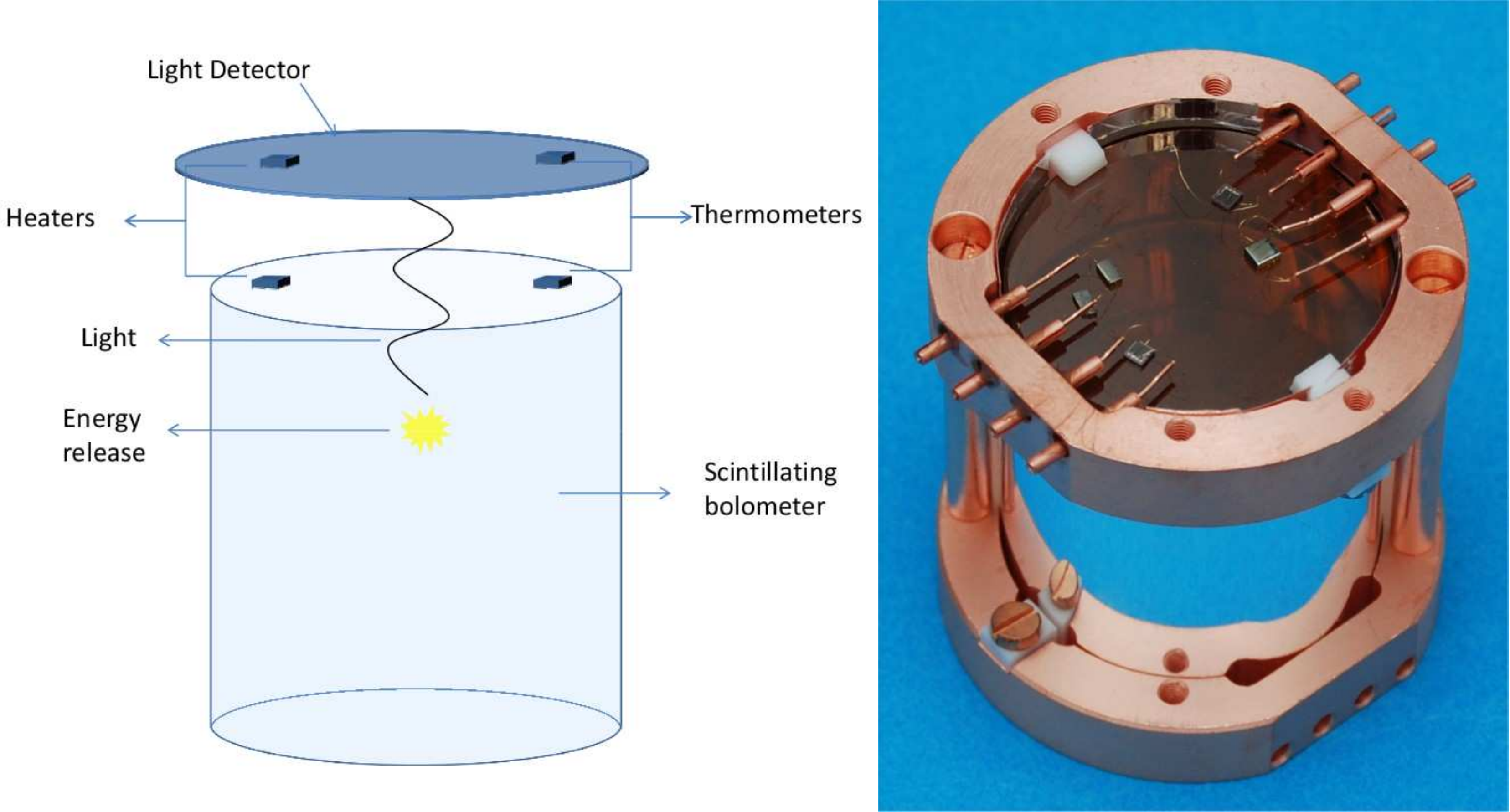}
\end{center}
\caption{Left panel: illustration of the operating principle of scintillating bolometers. The release of energy inside a scintillating crystal follows two channels: light production and thermal excitation. The heat is read out by a temperature sensor (NTD) glued on the primary crystal while the light is read by a second crystal (the light detector) where it is completely converted into heat. Right panel: the ZnSe Huge crystal in its Copper mounting structure. This crystal has few sensors glued on the top face, for redundancy. The light detector - here not visible - is mounted in his own Copper holder that fits in the top of the ZnSe one.}
\label{fig:bol_scint}
\end{figure}

The detector is in this case obtained by coupling a scintillating crystal - operated as bolometer - to a light detector (figure~\ref{fig:bol_scint}). When a particle traverses the scintillating crystal a (large) fraction of the deposited energy is converted into heat (following the chain described above), while a (small) fraction of it is spent to produce scintillation light (i.e. photons irradiated outside the crystal). For each interaction, two signals are recorded: a thermal signal produced in the scintillating bolometer and a photon signal read-out by the light detector.
The ratio between the two (heat vs. light) depends on the particle Light Yield (LY). Betas and gammas have the same LY, which is typically different from the LY of alphas or from that of neutrons. 
Consequently, the contemporary read out of the heat and light signals allows particle discrimination. 

In particular, if the scintillating crystal is made of a \BB candidate, the \BBz signal (i.e. the energy deposition produced by the two electrons emitted in the decay) can be distinguished from an alpha signal. This makes feasible the rejection of alpha induced background, opening a new window in the future of \BBz with bolometers \cite{Pirr06}. It is infact true that, in the recently concluded \BB bolometric experiment CUORICINO, the major source of background is identified in alpha particles due to surface contaminations \cite{ArtChambery,LNGSReport2006,paperFrank}. This same source could be the main factor limiting the sensitivity of the - presently under construction - CUORE experiment \cite{LNGSReport2006}. 

\section{$^{82}$Se \BBz decay}\label{sec:cryst}

$^{82}$Se is a \BB emitter with an isotopic abundance of 9.2\%  and a Q-value of (2995.5$\pm$2.7) keV \cite{Qvalue}. It has always been considered a good candidate for \BBz studies because of its high transition energy and the favorable nuclear factor of merit. 
Indeed, the more recent evaluations of the Se nuclear matrix element (NME), yield for this isotope a half-life in the range 1-8~10$^{26}$ y for an \amnu value of 50~meV~\cite{NME1,NME2,NME3,NME4} (in the same condition the predicted half-life for $^{76}$Ge is about 4 times larger).

In recent years the NEMO collaboration measured its half life for Two Neutrino Double Beta Decay (\BBdn) to be 9.6$\pm$0.3(stat)$\pm$1.0(syst)~10$^{19}$ y and set an upper bound on the \BBz half-life of 1.0~10$^{23}$ y at 90\% C.L. \cite{NEMO0n}.

In view of the realization of a high sensitivity \BBz experiment, the low isotopic abundance of  $^{82}$Se is a potential problem. Indeed, the experimental sensitivity depends on the number of \BB emitters. With natural Se a huge mass of material would be needed to have a reasonable number of \BB emitters. However, this problem can be solved by isotopic enrichment, as done - for this same isotope - by the NEMO collaboration \cite{NEMOdetector}. 

On the other hand, the high Q-value is an important factor of merit since places the \BBz signal in an energy region scarcely populated by the emissions due to natural radioactivity. 
This means that, in a calorimetric experiment, where the sum energy of the two electrons is recorded, a reduced number of sources contributes to the background and limits the experimental sensitivity \cite{Pirr06}. The measured \BBd half-life allows to evaluate the irreducible background due to this decay channel, that - for a calorimetric experiment with high resolution - results completely negligible (the background issue will be discussed in detail in section~\ref{sec:dbd}).

\section{Experimental details}

The results here discussed are a synthesis of an R\&D work performed over more than 2 years during which we have extensively studied the performances of ZnSe detectors in different conditions. For our studies we have used three different ZnSe crystals whose characteristics are reported in Table \ref{tab:crystals}. The crystals have been produced by Alkor Technologies (Saint-Petersbourg, Russia) and are undoped (usually ZnSe is doped with Te), their color variation is not fully understood, could be ascribed to a different stoichiometry \cite{artIoan}. 

ZnSe is a semiconductor with a Debye temperature of about 270 K \cite{Landolt-Bornstein} (not much different from \teodn), known as a luminescent crystal since many years \cite{luminescenza}. Its X-ray induced luminescence has been recently measured from room temperature down to 10 K, resulting in a dominant emission in the red region (610 and 645 nm) and a smaller component in the infrared region (970 nm) \cite{artIoan}. The Light Yield of this crystal (measured on small pieces cut from the same ingot from which the Small ZnSe crystal was obtained) was observed to increase at low temperature, with an emission characterized by more than one decay constant \cite{scintcresst}.
\begin{table}[h]
\caption{ZnSe crystal characteristics. In the last column the surface quality is indicated for the lateral surface and the two basis of the cylinder respectively. O stands for optical and R for rough.}
\begin{center}
\begin{tabular}{cccccc}
\hline\noalign{\smallskip}
crystal & crystal & mass & diameter & height & surface \\
name & coulor & [g] & [mm] & [mm] & quality \\
\noalign{\smallskip}\hline\noalign{\smallskip}
Small & yellow & 37.5 & 20 & 21 & O O O \\
Large & red & 120 & 41 & 17 & O R R\\
Huge & orange & 337 & 40 & 50 & O O O \\
\noalign{\smallskip}\hline
\end{tabular}
\end{center}
\label{tab:crystals}
\end{table}

In all the runs discussed here, the crystals were mounted in a Copper frame and tightly kept in position by PTFE tips. They were surrounded (without being in thermal contact) by a reflecting sheet (3M  VM2002), but for the side which is faced to the light detector (LD), as shown figure~\ref{fig:bol_scint}. The tests have been performed operating the ZnSe scintillating bolometers at about 15~mK, in a low temperature refrigerator installed underground in Hall C of Laboratori Nazionali del Gran Sasso (LNGS), L'Aquila - Italy. A group of more than 10 cryogenic runs, with a duration ranging from 1 month to few months, have been realized in order to come to the results described in this paper. In all these runs the bolometers worked properly, proving that this crystal bears thermal cycles without a deterioration of its performances, whereas it exhibits an abnormal cooling slowness.

Details of the set-up and the read-out are discussed in the next two sub-sections.

\subsection{ZnSe bolometers and the heat signal}\label{sec:boloheat}

Each ZnSe bolometer was equipped with a Neutron Transmutation Doped Ge thermistor (NTD) \cite{NTD}, glued on the crystal surface and used as a thermometer to measure the heat signal produced by particles traversing the ZnSe. 
A silicon resistance, glued on the crystals, was used to produce a calibrated heat pulse in order to monitor the thermal gain of the bolometer. This is indeed subject to variation upon temperature drifts of the cryostat that can spoil the energy resolution. In most cases this temperature drift can be re-corrected off-line on the basis of the measured thermal gain variation~\cite{articoloimpulsatore,articolostabilizzazione}.  

The NTD thermistor transduces the heat pulse into a voltage signal which is amplified and fed into a 16 bit ADC. The entire waveform (\emph{raw pulse}) is sampled, digitized and acquired. At the ADC input, a low pass Bessel filter is used as antialiasing to avoid spurious contributions in the sampled signal \cite{elettronica-OF}. 

The amplitude V$^{heat}$ and the shape of the voltage pulse is then determined by the off-line analysis that makes use of the Optimal Filter technique \cite{elettronica-OF,GattiManfredi}. The signal amplitudes are computed as the maximum of the optimally filtered pulse, while the signal shape is evaluated on the basis of four different parameters: $\tau_{rise}$ and $\tau_{decay}$, TVL and TVR. 

$\tau_{rise}$ (the rise time) and $\tau_{decay}$ (the decay time) are evaluated on the \emph{raw pulse} as (t$_{90\%}$-t$_{10\%}$) and (t$_{30\%}$-t$_{90\%}$) respectively. For these detectors they are of the order of 5-10 ms and 20-30 ms. The rise time is mainly dominated by the time constant of the absorber-glue-sensor interface, while the decay time is determined by the crystal heat capacity and by its thermal conductance toward the heat sink. 

TVR (Test Value Right) and TVL (Test Value Left) are computed on the optimally filtered pulse as the least square differences with respect to the filtered response function\footnote{The response function of the detector, i.e. the shape of a pulse in absence of noise, is computed with a proper average of a large number of raw pulses. It is also used, together with the measured noise power spectrum, to construct the transfer function of the Optimal Filter.} of the detector: TVR on the right and TVL on the left side of the optimally filtered pulse maximum. These two parameters do not have a direct physical meaning, however they are very much sensitive (even in noisy conditions) to any difference between the shape of the analyzed pulse and the response function. Consequently, they are used either to reject fake triggered signals (e.g. spikes) or to identify variations in the pulse shape with respect to the response function (and this will be our case, as in section~\ref{sec:PSA}). \\

\begin{figure}[]
\begin{center}
\includegraphics[width=1\linewidth]{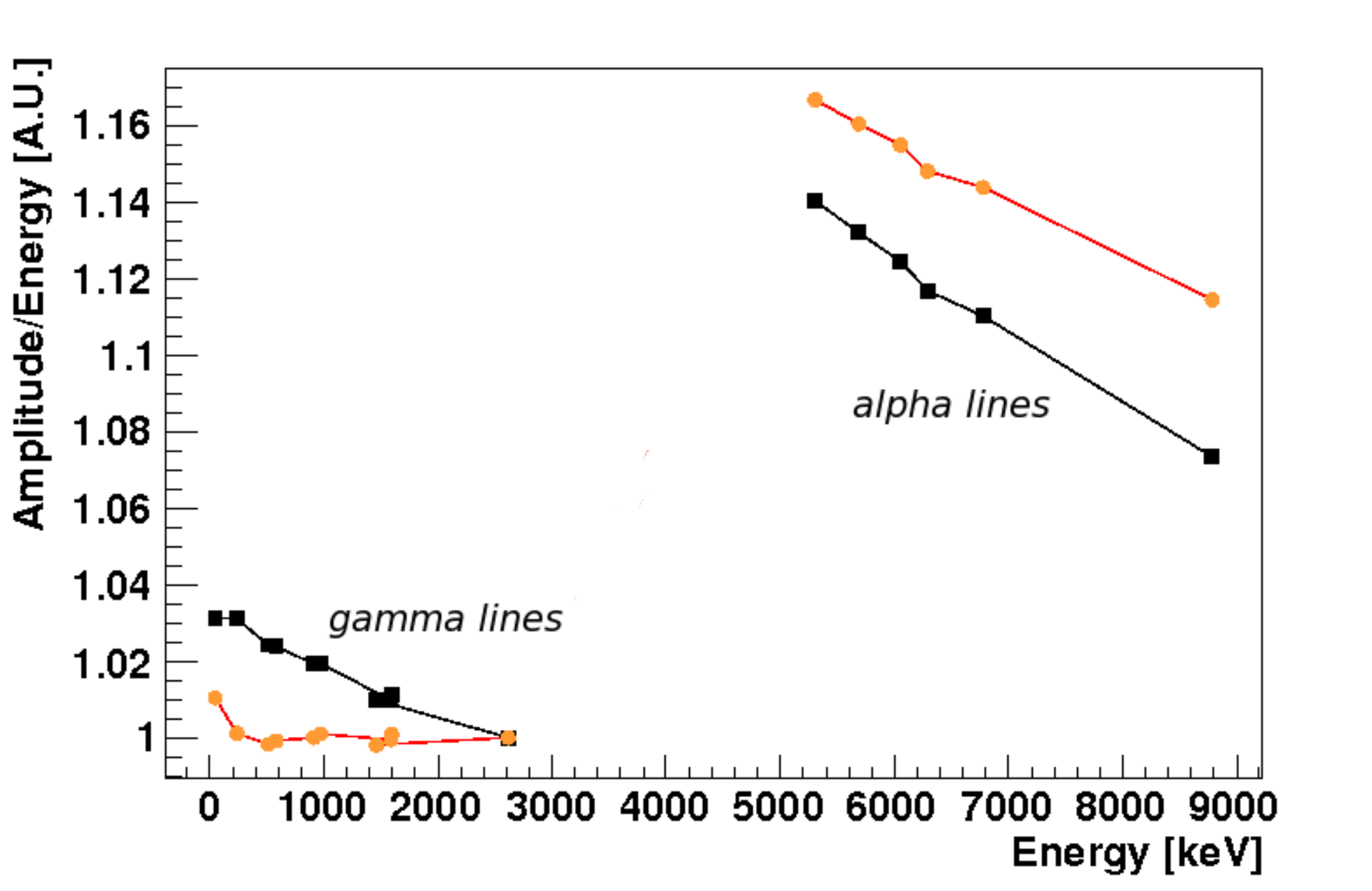}
\end{center}
\caption{Non linearity behaviour of the ZnSe bolometer. On the X axis the nominal energy ($E$) of gamma (\thdt and \kqn) and  alpha (source A) lines. On the Y axis the ratio between V$^{heat}$ and $E$ (black squares) before the energy linearization of the bolometer response (the ratio is normalized at 1 for the 2615~keV peak). For the same lines the ratio between $Heat$ and $E$ is reported (colored circles). These data refer to the Large ZnSe crystal. A similar behaviour is observed for all the other detectors: the gamma pulses show a non linearity of the order of 3\% on a 3 MeV range, the alpha lines have a similar non linearity but a completely different energy calibration. The experimental errors are of the order of 0.1-0.2\% and are not visible in the plot.}
\label{fig:heatnonlinearity}
\end{figure}

In each run, the ZnSe bolometers were calibrated with a \thdt source placed between the cryostat and its external lead shield. Gamma lines between 238~keV and 2615~keV were used for amplitude (V$^{heat}$) vs. energy ($E$) linearization of the heat signal. The response of the bolometer is indeed slightly non linear, mainly because of the exponential dependence of the NTD thermistor resistance from temperature. In figure~\ref{fig:heatnonlinearity} we report the ratio V$^{heat}$/$E$ vs. $E$ (where $E$ is the full energy of the corresponding gamma peak) observed for one of our detectors. The points with $E$\less 3 MeV correspond to gamma peaks identified in the heat signal spectrum, the others correspond to alpha lines (these are due to a $^{224}$Ra source facing the detector, see section~\ref{sec:results} and figure~\ref{fig:sourceA} for more details). The non linearity measured on gamma lines is here of about 3\% on a 3~MeV energy range. Non linearities of the same order of magnitude are observed for all the other detectors. As it is usually done for other bolometers \cite{paperFrank,paperocdwo4}, the non-linearity is corrected by introducing the calibrated amplitude of the heat pulse ($Heat$) - obtained from V$^{heat}$ - using the following relationship:
\begin{equation}
log(Heat)= A+B\,\,log(V_{heat})+ C\,\,log(V_{heat})^2
\end{equation}
where the three coefficients A, B and C are determined fitting the (V$^{heat}_i$, $E_i$) points of the calibration gamma lines. 
It is worth to note that in this way we attribute to the heat signal the \emph{nominal} energy $E$ of the detected gamma line, i.e. the calibration returns $Heat$=$E$ for gamma particles. Here $E$ is the total energy lost within the crystal, while only a fraction of it is converted into heat. The other fraction is spent in scintillation light or in possible \emph{blind} channels such as long living metastable states. 
The result obtained after the calibration is shown in figure~\ref{fig:heatnonlinearity}: the colored circles are the ratio $Heat$/$E$ for each gamma and alpha line. The calibration returns, as expected, $Heat$/$E$\ca 1 for all the gamma lines. While for alpha lines $Heat$/$E>$1 (all the points with $E$ $>$ 3~MeV in figure~\ref{fig:heatnonlinearity}). This means that alpha lines appears in the $Heat$ spectrum with a displacement of about 1 MeV, i.e. for the same total energy deposition in the crystal, heat pulses due to alphas are higher (by a factor \ca1.15) than heat pulses due to gammas. Provided that we observe for alpha particles a higher light signal, this behaviour is quite strange. We will come back to this feature in section~\ref{sec:LY}.\\\\

\begin{figure}[]
\begin{center}
\includegraphics[width=1\linewidth]{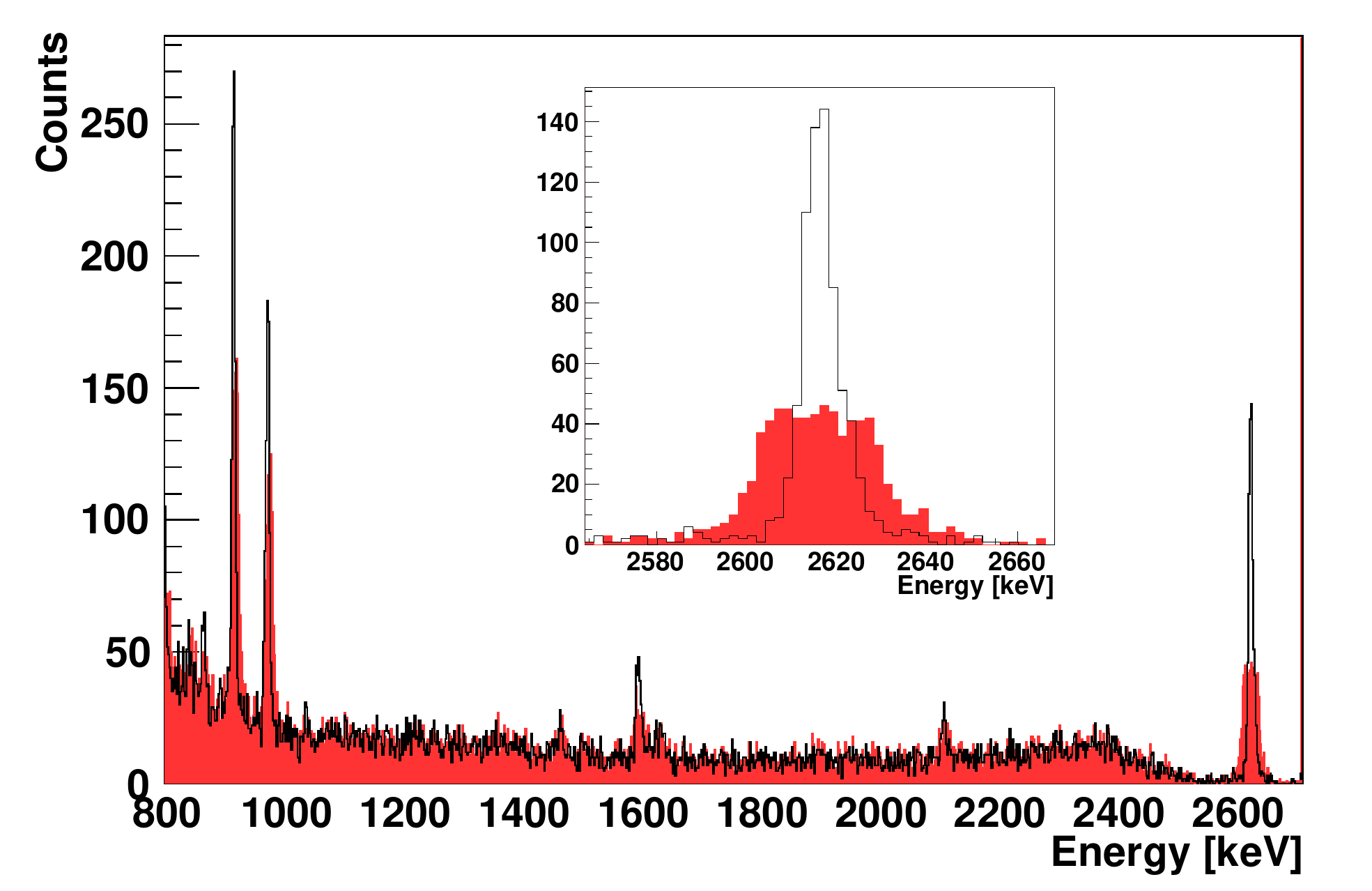}
\end{center}
\caption{\thdt gamma calibration of the Huge ZnSe crystal. The filled histogram is the projection - on the $Heat$ axis - of the events visible in the scatter plot of figure~\ref{fig:scatthuge} (right panel).  The black line histogram is obtained as projection on the $Heat_{\theta}$ axis (where $\theta$ has the value that optimize the RMS of the 2615~keV \tld peak), see section~\ref{sec:correl} for more details. In the inset the highlight of the 2615~keV line. The gain in resolution on all the lines is evident.}
\label{fig:rotazione}
\end{figure}

The energy resolution on the heat signal was evaluated measuring the FWHM of the 2615~keV line in the \emph{Heat} spectrum. The peak FWHM was evaluated by a fit with an asymmetric gaussian. The best resolutions measured (average of left and right FWHM) for the Large and Huge crystals were of 21$\pm$2~keV and 28$\pm$1~keV respectively (see filled histogram in figure~\ref{fig:rotazione}). 

In the case of the ZnSe Small crystal we have not been able to evaluate the energy resolution because no gamma peak is visible in the calibration or background spectra. This is explained by the small dimensions of the crystal and its low Z. Since we only used it at the beginning of our studies on ZnSe, we did not have the occasion for a long calibration measurement.
 
\subsection{Light detectors and the photon signal}\label{sec:bololight}

The scintillation light emitted by the ZnSe crystals is detected with small bolometers \cite{NIMA-2006}. These are made by high purity Ge wafers with a large area (to optimize photon collection) but a small mass (to reduce the heat capacity and therefore to have a sizable temperature increase upon photon absorption). They are 36~mm in diameter and 1~mm in thickness. The surface of the Ge wafers facing the scintillating crystal are \emph{darkened} through the deposition of a 600~\AA\ layer of SiO$_2$ in order to increase the light absorption.
We have used also a very large area Ge wafer (66~mm in diameter) able to cover contemporary the Small and Large ZnSe crystals in order to provide a direct comparison of their Light Yields. 

The Ge wafers are provided, as the ZnSe bolometer, with an NTD thermistor for the temperature read-out and with a Si resistor for gain monitoring. Electronics and DAQ are identical to that of ZnSe scintillators.  Rise and decay times of the signal are of about 2 and 15~ms respectively.

The Ge wafer are too small to be able to detect the energetic gamma lines of the \thdt source used for ZnSe calibration. 
Therefore a weak $^{55}$Fe source was faced to the Ge wafers, on the side opposite to the crystal scintillator, for calibration purposes. The 5.9 keV K$_{\alpha}$ X ray line of $^{55}$Fe was used to measure the energy resolution of the detector and its amplitude to energy conversion. Obviously, with only one calibration peak, we are not able to correct the $Light$ amplitude for the non linearity of the LD response. However, the expected non linearity is similar to the one recorded for the scintillating bolometer (since the NTD sensor is similar) while the energy range is smaller by more than one order of magnitude. The effect should be therefore negligible.

The $^{55}$Fe calibration provides a method to measure and compare the LY's of different crystals. This cannot be used for an absolute evaluation of the amount of scintillation light emitted by the crystal since the $Light$ amplitude is not corrected for the photon collection efficiency that is certainly below 1.
Finally the energy resolution of the LD is measured as the FWHM of the 5.9 keV line. Depending on the detector used and on the working conditions, the resolutions range from 200~eV to 600~eV FWHM. These resolutions are negligible if compared with the width of gamma and alpha lines in the $Light$ spectrum. 

\begin{figure}
\begin{center}
\includegraphics[width=1\linewidth]{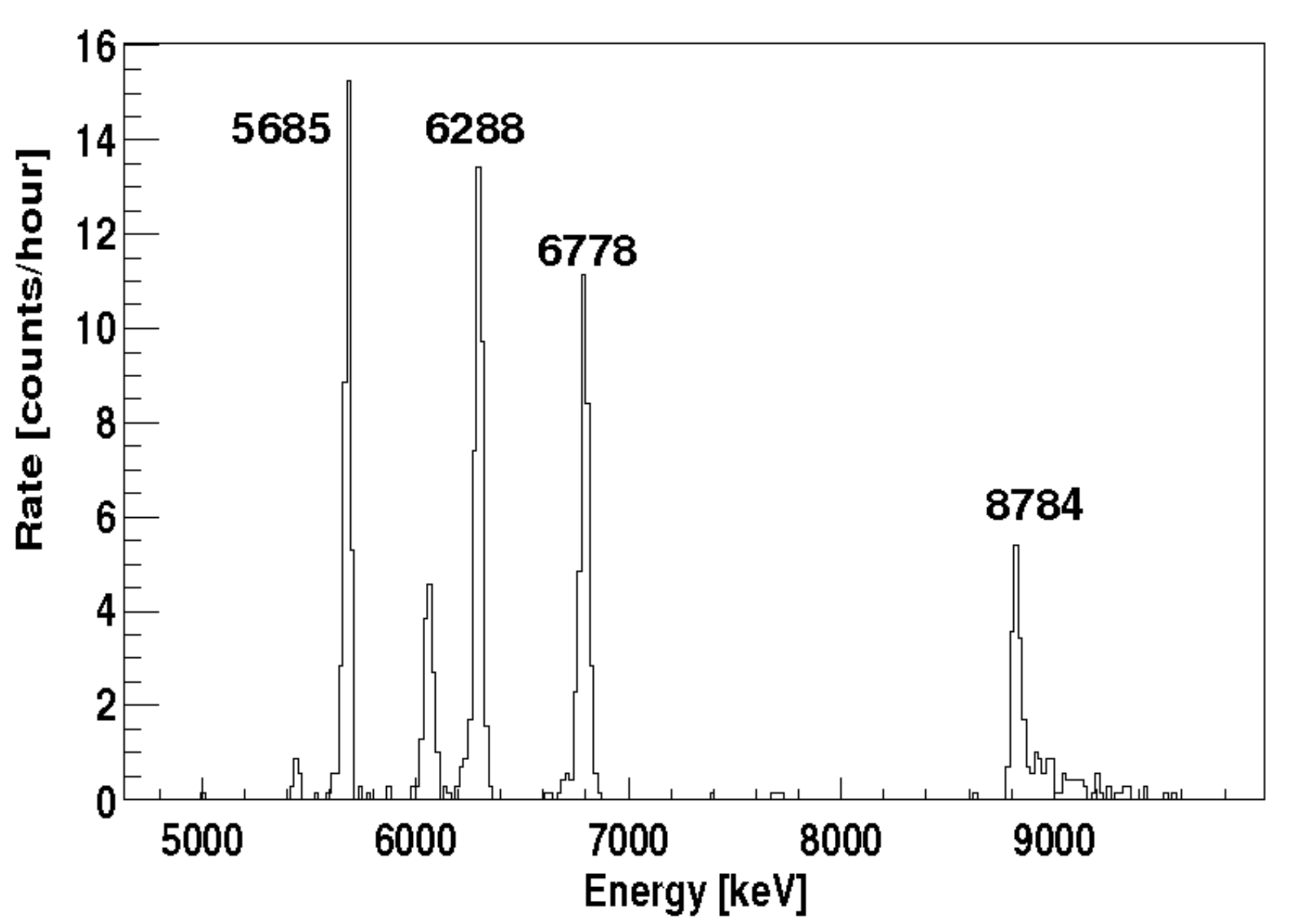}
\end{center}
\caption{Spectrum of source A as recorded by a Si surface barrier detector (having 35 keV FWHM energy resolution at 5 MeV). The energy [keV] of the main lines is indicated.}
\label{fig:sourceA}
\end{figure}

\section{Results}\label{sec:results}

We discuss in this section the results obtained in several runs performed in the last few years, with the three above mentioned crystals. Our measurements aimed mainly at the study of the alpha vs. beta/gamma discrimination properties of ZnSe, at the evaluation of the LY and of the Quenching Factor (QF), and at the study of the variation of these properties upon variation of the crystal color and surface quality. 
In order to be able to compare the detector response to alpha and beta/gamma particles, in most of the runs, the crystals were faced to low intensity alpha sources (while gammas were generated with the external \thdt source used for calibration).
Two different radio-isotopes have been used so far.

Source A was obtained by implantation of $^{224}$Ra in an Al reflecting stripe. The implantation was accomplished by exposing the stripe to a thin $^{228}$Ra source. The $^{228}$Th isotopes contained in the source alpha decay to $^{224}$Ra. The recoil energy of $^{224}$Ra is sometimes enough to eject the nucleus from the source and implant it in the Al stripe. The extremely short range of nuclear recoils in solid materials allows to obtain a very shallow depth source (this can be appreciated in figure~\ref{fig:sourceA} from the symmetry of the peaks). Besides several alpha lines of different relative intensity, this source also produce beta particles with a maximum energy of 5 MeV due to the beta decay of \tldn.

Source B was obtained contaminating an Al stripe with a \udt liquid solution and later covering the stripe with an alluminated Mylar film (6 $\mu$m thick). Thus the source has a continuous spectrum extending from about 3 MeV to 0.\\

The usual way to visualize alpha vs. beta discrimination is to draw the $Light$ vs. $Heat$ scatter plot. 
Here each event is identified by a point whose abscissa is equal to the $Heat$ signal amplitude (recorded by the scintillating bolometer), and ordinate equal to the $Light$ signal amplitude (simultaneously recorded by the LD). In the scatter plot beta/gamma, alphas and neutrons give rise to separate bands, in virtue of their characteristic $Light$ to $Heat$ ratio. As already anticipated, this feature is the result of the different LY's characterizing these particles. 

\begin{figure*}
\begin{center}
\includegraphics[width=1\linewidth]{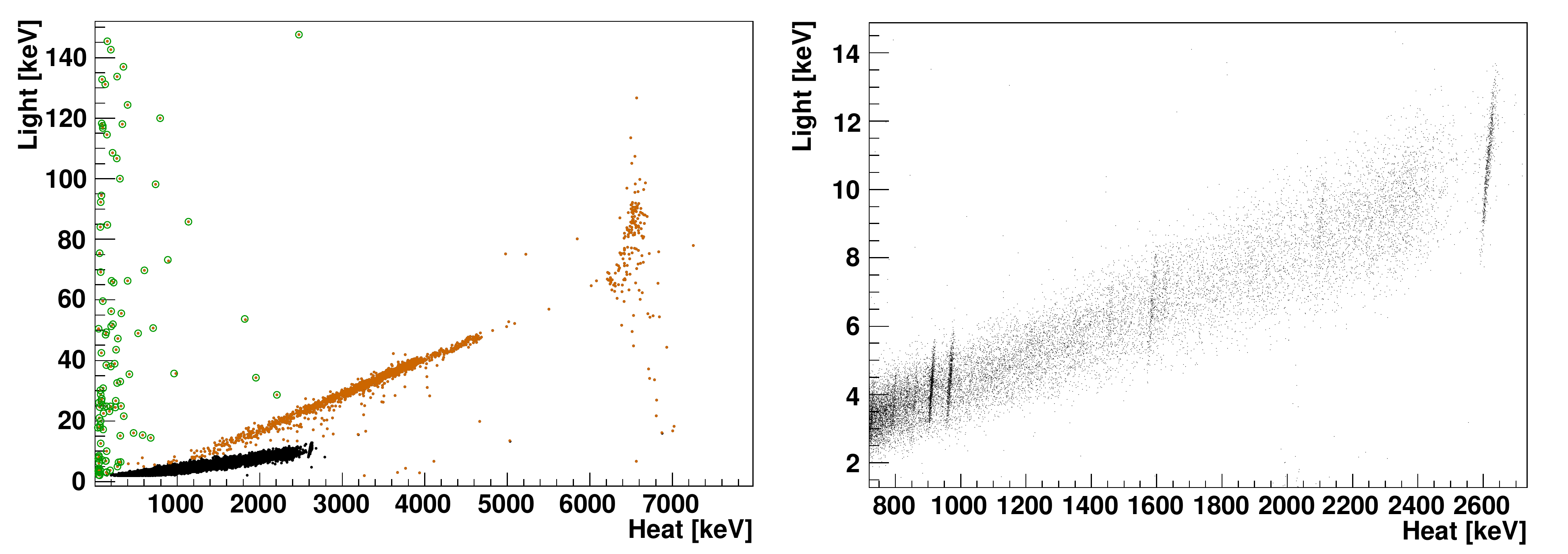}
\end{center}
\caption{Scatter plot $Light$ vs. $Heat$ recorded with the Huge ZnSe crystal. The detector was here exposed to the B alpha source and to the calibration gamma source ($^{232}$Th located outside the refrigerator). Three populations of events have been marked: black spots for beta/gamma's, colored spots for alphas, circles for events with an ionization component in the LD. In the right panel a zoom of the beta/gamma band. The gamma lines due to \thdt are clearly visible. The slope of all the lines is positive.}
\label{fig:scatthuge}
\end{figure*}

The $Light$ vs. $Heat$ scatter plot of the ZnSe Huge crystal, exposed to the calibration gamma source (\thdtn) and to the alpha source B, is shown in figure~\ref{fig:scatthuge}.  This scatter plot (as well as all the others we obtained with ZnSe crystals) is completely different from any other we ever measured with different scintillating crystals (we tested several scintillating bolometers like CdWO$_4$, ZnMoO$_4$ and CaMoO$_4$). Three anomalies are indeed here evident:

\begin{enumerate}
\item monocromatich gamma peaks lie on positive slope lines (right panel of figure~\ref{fig:scatthuge}), whereas a negative slope is expected;
\item the alpha band lies above the beta/gamma band, suggesting an alpha QF larger than one (left panel of figure~\ref{fig:scatthuge});
\item alphas are always accompanied with a tail that draws negative slope lines (figure~\ref{fig:codealfa}).
\end{enumerate}

 The anomalies resulted reproducible for the three crystals, despite the different crystal colour and size and despite the different experimental conditions. In the following three sections we document these unexpected features and try to give a possible interpretation of them.

\subsection{Correlation between $Heat$ and $Light$ and energy resolution on gammas}\label{sec:correl}

In a \emph{standard} scintillator, monochromatic events (i.e. those corresponding to the complete absorption of a monochromatic particle in the scintillating crystal) appear as sections of lines with negative slope, showing an anticorrelation between $Heat$ and $Light$ \cite{paperocdwo4,coron}. The observed spread along a negative slope line can be accounted for assuming the existence of fluctuations in the $Heat$ and $Light$ signal amplitudes. Since the energy has to be conserved, the amplitude of the fluctuations ($\delta Light$ and $\delta Heat$) must compensate each other: $\delta Light$=-$\delta Heat$. In other words the fluctuations are anti-correlated between each other and produce the observed pattern. Finally, they are visible because their amplitude is much larger than the intrinsic resolution of the detectors. This is what happens for example for CdWO$_4$ crystals. In this case a rotation of the axis can be used to improve the energy resolution of the detector \cite{paperocdwo4,coron}.

In ZnSe crystals the slope of gamma lines is clearly positive (see right panel of figure~\ref{fig:scatthuge}): for a monochromatic event (as it is the photopeak of a gamma line) the minimum heat release corresponds to the minimum of light emission, and vice-versa. It seems that fluctuations in the energy conversion into heat and light are correlated, and that there is a lacking energy which is feeding some invisible channel. 

The data presently in our hands are not enough to proceed further in the modeling of this  unusual behaviour of ZnSe scintillator, we simply observe that - also in this configuration - a rotation of the axis \cite{paperocdwo4} can be used to improve the energy resolution of the detector.

Figure ~\ref{fig:rotazione} shows (filled histogram) the spectrum obtained projecting the gamma events of figure~\ref{fig:scatthuge} on the $Heat$ axis. The black line histogram is obtained by projecting events on a rotated axis ($Heat_{\theta}$) defined as:

\begin{equation}
Heat_{\theta}=Heat\,cos(\theta)-Light\, sin(\theta)
\end{equation}

\noindent where the rotation angle $\theta$ is chosen in order to reach the minimum FWHM for the 2615~keV. The improvement obtained on the resolution is of about a factor 3 on the 2615~keV line of \tld (the FWHM is reduced from 28$\pm$1~keV to 9.5$\pm$0.4~keV). On the other peaks the improvement is smaller: at 911~keV from 7.8$\pm$0.6~keV to 5.9$\pm$0.3~keV, at 583 keV from 7.5$\pm$0.5~keV to 5.4$\pm$0.3~keV. Provided that it is proved that also \BBz events would give rise to a monochromatic line with the same slope, this technique could be used to improve the resolution of the detector.

\subsection{Light Yield and anomalous light emission from the alphas}\label{sec:LY}

 The Light Yield of a crystal is conventionally defined as the fraction of the total energy deposited in the scintillator which is converted into light. As discussed previously, we do not have an absolute calibration of the $Light$ signal, therefore we can provide only a relative evaluation of the Light Yield defined as:
 \begin{equation}
LY_{\gamma}=  \frac {Light_{\gamma}} {Energy_{\gamma}}
\end{equation}
A similar definition holds for alphas. In our case, we chose to evaluate the LY$_{\gamma}$ on the 2615~keV line of \tld (i.e. Energy$_{\gamma}$=2.615~keV). The results for our 3 crystals are reported in table \ref{tab:ly}. 

\begin{table}
\begin{tabular}{cccc}
\hline\noalign{\smallskip}
Crystal & ZnSe Small & ZnSe Large & ZnSe Huge\\
\noalign{\smallskip}\hline\noalign{\smallskip}
LY [keV/MeV]& 1.3 & 7.5 & 4.6 \\
QF$_{\alpha}$ & 4.4 & 4.2 & 3  \\
\noalign{\smallskip}\hline
\end{tabular}
\caption{LY$_{\gamma}$ measured on the 2615~keV line of \tld (\thdtn) and QF$_{\alpha}$ measured for the 5.7 MeV alpha line of $^{224}$Ra. For each crystal we report values measured in similar experimental conditions: with the LD mounted on the top of the crystal and the alpha source facing the opposite side. The experimental errors on LY$_{\gamma}$ and QF$_{\alpha}$ are of the order of 5\% and 10\% respectively. In the case of the ZnSe Small and Large crystals we were also able to compare directly their LY$_{\gamma}$ in a dedicated run, where their scintillation photons were detected by the same Ge wafer (6~mm in diameter). The results where consistent with what reported in this table.}
\label{tab:ly}
\end{table}
 
 The QF$_{\alpha}$ is usually defined as the ratio between the amplitudes of the light pulses corresponding to alpha and beta/gamma particles of the same energies. It is however difficult to have alpha and beta/gamma sources with monochromatic lines of the same energies therefore we have evaluated the QF$_{\alpha}$ as:
\begin{equation}
QF_{\alpha}=  \frac {Light_{\alpha}} {Energy_{\alpha}} \cdot \frac {Energy_{\gamma}} {Light_{\gamma}}
\end{equation}
where we always compare the 5.7 MeV alpha line of the implanted source with the 2615~keV gamma line of the \thdt source.

The QF obtained for the Small, Large and Huge crystals, for an identical position of the alpha source (i.e. facing the crystal basis opposite to the LD) are reported in table \ref{tab:ly}.
QF$_{\alpha}$ values are definitely larger than 1, and this unusual feature is accompanied by other peculiar behaviours that we extensively investigated in several dedicated runs. The experimental results and the conclusions that can be drawn are discussed in the next subsections.


\subsubsection{The anomalous alpha QF}
To our knowledge no other scintillator with a QF$_{\alpha}$ definitely higher than 1 have ever been reported in literature, and this feature cannot be easily accommodated within the theoretical framework used to describe the scintillation properties of materials. Indeed heavy particles usually have QF lower than one and this is explained on the basis of a saturation effect due to the high ionization density that characterizes the interaction of heavy particle in matter \cite{Birks}.

We have considered the possibility that a QF$_{\alpha}$ larger than 1 could be due to the experimental configuration, in some way privileging the collection of photons emitted by the point-like external alpha source with respect to the light produced in crystal bulk by gamma rays. We investigated therefore the light collection, the self-absorption of the scintillation photons inside the crystal and - finally -  the transparency of the LD's to the scintillation photons.
In more detail:
\begin{itemize}
\item a self-absorption of the light inside the scintillating crystal could penalize gamma interactions (that take place in crystal bulk) with respect to alpha ones (alphas come from external sources and interact in a thin surface layer). To investigate this effect we placed the alpha source on the side opposite to the LD in such a way that only the light traversing the crystal could be detected. All the QF's reported in table \ref{tab:ly} have been measured in this configuration. 
\item to check possible problems arising from light collection we compared a measurement done with a ZnSe Huge crystal and a CdWO$_4$ crystal of the same size. Both crystals were mounted in a completely identical configuration with the same LD. 
The result was a \emph{normal} behaviour in the CdWO$_4$ crystal and an \emph{anomalous} one in the ZnSe. The LY$_\gamma$ measured for the CdWO$_4$ crystal was 17 keV/MeV \cite{paperocdwo4}. Consequently, ZnSe has a light emission on gammas that is four times lower than the one recorded with CdWO$_4$, while the light emission on alphas is nearly two times larger.
\item we investigated the possibility that the low LY$_\gamma$ we measure could be due to a dominant beta/gamma photon emission on a wavelength to which our LD is partially transparent. The Ge wafers are completely opaque to the red light that, as measured upon X-rays excitation \cite{artIoan}, appears to dominate the ZnSe luminescence spectra. Nonetheless, at the very low temperature where we work, something could change. To verify this hypotesis we made a test using - as LD - a sapphire bolometer covered with a lead substrate (completely opaque from IR to UV). No relevant modification in the measured LY's and QF's was observed.
\end{itemize}
All our measurements, and consequently the above reported considerations, have been based on external alpha sources since we have never observed - in our crystals - an internal alpha line. We investigated the possibility that bulk alphas do not scintillate, in this case we should see heat signals in anticoincidence with light ones. These \emph{dark} counts are present in our spectra, but with the same percentage in the alpha and gamma region. They can in fact be ascribed to a light-heat coincidence lost due to trigger inefficiency on one of the two channels.

\begin{figure}
\begin{center}
\includegraphics[width=1\linewidth]{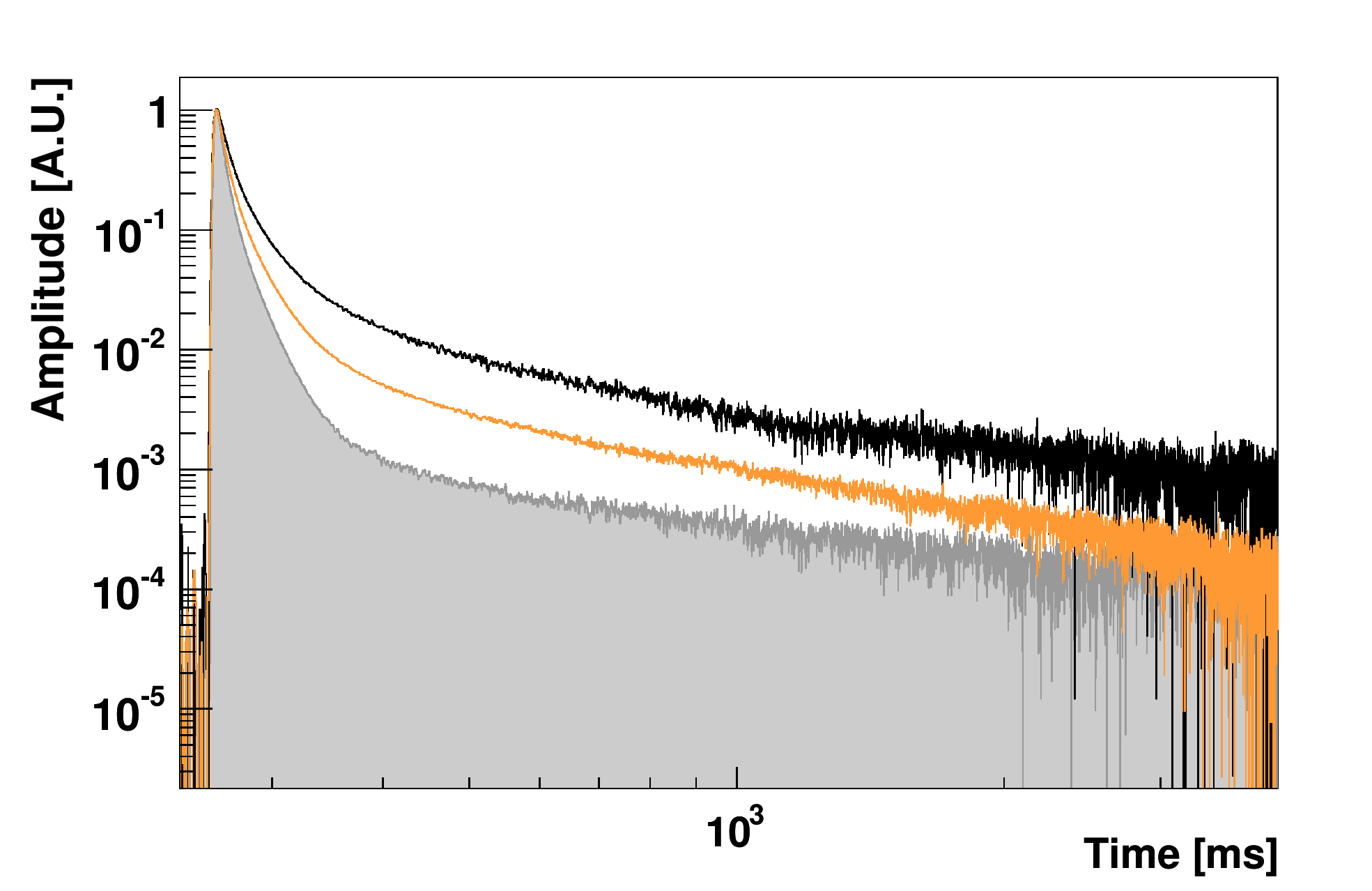}
\end{center}
\caption{Averaged light pulses from beta/gamma (black line) and alpha (colored line), after pulse height normalization. In shadow the response function of the LD, as obtained averaging pulses due to direct ionization in the Ge wafer. The real shape of the light signals shall be obtained by deconvolving the signal with the response function of the detector. The time window is 4~sec. wide.}
\label{fig:lightpulse}
\end{figure}

As a final remark on this topic, it is worth to recall that the value of the QF$_\alpha$ may depend on the parameter that is used to measure the energy content of the recorded pulse. In our case this is the pulse amplitude. In a very naive model of the bolometer, that assumes an instantaneous energy deposition in a monolithic device \cite{mather}, the amplitude is proportional to the energy released in the crystal. However, we have clear indications that these approximations do not hold. As it will be discussed in section~\ref{sec:PSA}, we observe a clear difference between the shape of alpha and beta/gamma light pulses. 
Figure~\ref{fig:lightpulse} shows averaged alpha and beta/gamma light pulses having the same $Heat$ amplitude (and consequently produced by particles releasing in the scintillator almost the same energy).\footnote{The time window is here 4~sec. wide. The average was needed to reduce the noise contribution and clearly appreciate the long tail of the pulses. It has to be remarked that - in the amplitude interval where we averaged pulses - no dependence of pulse shape on amplitude is observed (i.e. the average does not change the shape of the signal).} Once normalized to the same $Light$ amplitude the two signals show a quite different area. If the area, and not the amplitude of the pulse, would be used as an estimator of its energy content the QF$_{\alpha}$ would result lower than reported in table \ref{tab:ly} (by about a factor 1.5), although still much higher than~1. It could therefore be argued that, upon collection of all the light emitted by beta/gammas, the QF$_{\alpha}$ could approach a more \emph{standard} value.

For the sake of completeness, we mention here that a strange behaviour of the QF$_{\alpha}$ in ZnSe:Te (Tellurium doped Zinc Selenide) was already reported in literature. 
In particular, E.~V.~Sysoeva et~al.~\cite{Sysoeva} quote a QF$_{\alpha}$=1.05. While W. Klamra et al. \cite{Klamra} quote  a QF$_{\alpha}$=0.74-0.69, depending on the shaping factor of the amplifier. 
It is however very difficult to compare our results with those reported in literature.  There are indeed two key aspects that have to be pointed out:             
\begin{itemize}
\item all the measurement performed before this work were carried on extremely thin crystals (1-2 mm). This because the transmission band of ZnSe - at room temperature - is very close to the emission band (while at low temperatures the crystal becomes almost transparent at the 610~nm wavelength of the dominant emitted radiation).
\item our measurements are done on undoped ZnSe crystals at extremely low temperatures, while most of the literature concentrates on doped samples measured at much higher temperatures. In our case the choice is motivated by the fact that the doping introduces a strong modification of the crystal lattice, resulting in a larger heat capacitance, thus strongly decreasing the performances of a bolometer. On the other hand the use of an undoped crystal is advantageous because - as well established - decreasing the temperature the LY of intrinsic scintillators increases, while the opposite happens for doped scintillators.
\end{itemize}

\subsubsection{QF dependence on source position}

\begin{figure}
\begin{center}
\includegraphics[width=1\linewidth]{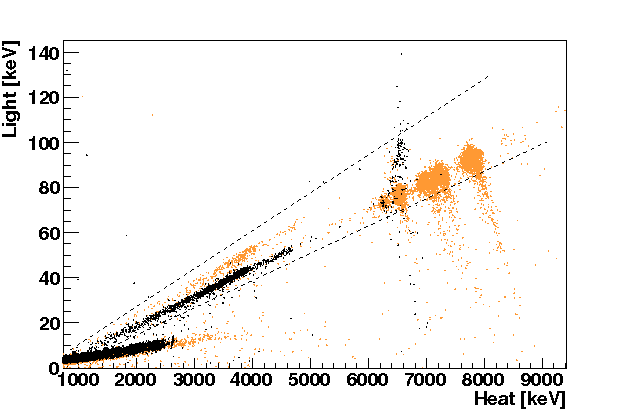}
\end{center}
\caption{Scatter plot of $Light$ vs. $Heat$ recorded with the Huge ZnSe crystal in two different runs. In run I (colored points) the B alpha source was mounted on the bottom side of the crystal, opposite to the LD. The scatter plot obtained in this run is reported in figure~\ref{fig:scatthuge}. In run II (black points) source B was moved on the side of the crystal and in its place was mounted source A. The dotted lines identify the \emph{opening} of the alpha band on a wide angle, due to the variation of the QF$_{\alpha}$ upon source position.}
\label{fig:scatthuge2sources}
\end{figure}

Figure~\ref{fig:scatthuge2sources} compares the results obtained with the same detector changing only the position of the alpha sources. In run I the B alpha source was mounted on the bottom side of the crystal, opposite to the LD (this is the same scatter plot reported in figure~\ref{fig:scatthuge}). In run II source B was moved on the side of the crystal, while in its previous position was mounted source A. By comparing the QF$_\alpha$ obtained for the two positions, it is clearly evident a variation of the order of 20\% . This indicates that - in the case of a source distributed uniformly around the crystal (and not point-like sources like those we are using)- we should observe an alpha band spanning a much larger angle.

In figure~\ref{fig:scatthuge} and figure~\ref{fig:scatthuge2sources} we observe the presence of background alpha events forming a positive slope line (similar to the ones observed for gamma particles, see section~\ref{sec:correl}) at $Heat$\ca 6.2~MeV and $Light$\ca80~keV. This region corresponds to a nominal alpha energy of about 5.3~MeV, which is the energy of the alpha decay of \poddn. Indeed, as discussed in section~\ref{sec:boloheat}, the correct energy of an alpha particle is lower than the recorded $Heat$ amplitude by about a factor 1.15. This line can be ascribed to a \pbdd contamination (a quite common contaminant whose strongest signature is the alpha emitted by \podd) because its intensity does not change in time (\podd has a half-life of 138~days and we have two measurements with the same crystal at a distance of \ca 4 months, without sizable variation of intensity). The line has an evident tail (with negative slope) toward low $Light$ - high $Heat$ signals, an indication (as will be discussed in section~\ref{sec:alphatail}) that it is due to a surface contamination. A reasonable assumption is that the source is a \pbdd surface contamination of the crystal or of the mounting structure surrounding the crystal. In this case, the extension of the line on the $Light$ axis can be interpreted as due to the variation of the QF$_{\alpha}$ as a consequence of a position dependence. In figure~\ref{fig:scatthuge2sources} we have indicated approximately the width that we could expect for the alpha band assuming that the \podd line is due to an uniform contamination surrounding the crystal.

The dependence of QF$_{\alpha}$ on the source position could in principle be due to a difference in photon collection efficiency, however in a test run done with two LD (where therefore the collection efficiency was definitely improved) the extension of the \pbdd line on the $Light$ axis was not observed to change. Moreover, upon changes in the alpha source position, we observe a variation not only in QF$_{\alpha}$ but also in pulse shape (the decay time of the light signals). All this gives credit to a second hypothesis: that the variation of QF$_{\alpha}$ upon source position is due to position dependence of LY$_{\alpha}$ (a similar effect is discussed in \cite{Danevich}) or to the  self-absorption of the emitted light.

\subsubsection{Low Light tails of the alphas in the scatter plot}\label{sec:alphatail}
\begin{figure}
\begin{center}
\includegraphics[width=1\linewidth]{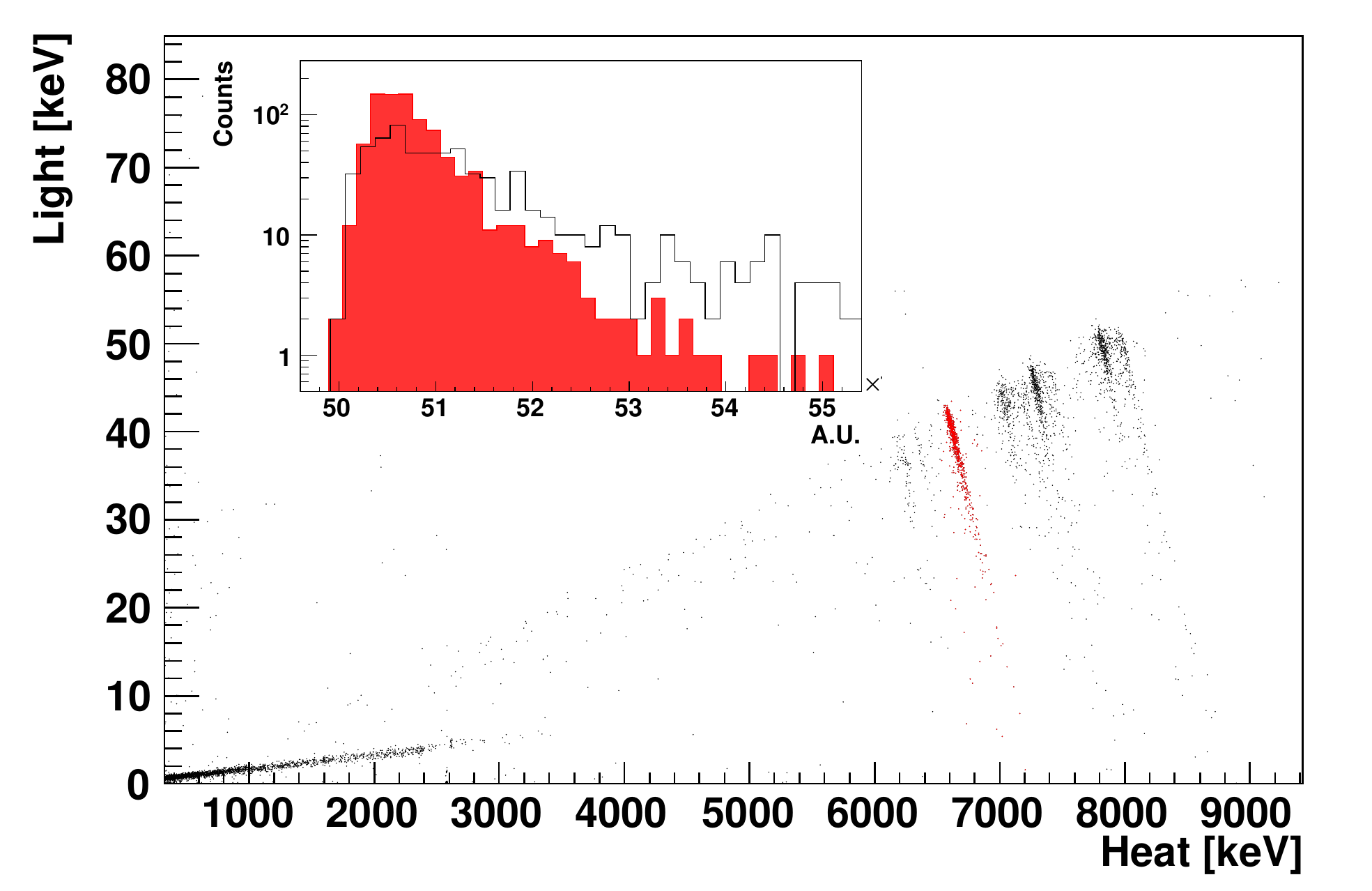}
\end{center}
\caption{Scatter plot of $Light$ vs. $Heat$ recorded with the Large ZnSe crystal. The detector was here exposed to the A alpha source and to the calibration gamma source (\thdt located outside the refrigerator). Each alpha peak comes with a - low $Light$ high $Heat$ - tail that draws a negative slope line in the scatter plot. In the inset: histogram of the events belonging to the 5.7 MeV alpha line (colored spots) projected along their negative slope line. The two histograms have been obtained for the same source (source A) and the same detector. The only difference being that in the case of the black histogram the non optical basis of the ZnSe cylinder was in front of the source, while in the case of the colored histogram it was the optical basis that faced the source (the scatter plot refers to this latter configuration). }
\label{fig:codealfa} 
\end{figure}

As anticipated in the previous section, the separation of the alpha band from the beta/gamma one is not complete. This is evident looking at monochromatic alpha lines, like those produced by source A.  For each alpha line, we observe a spot plus a long tail of events - with lower $Light$ and larger $Heat$ - that draws a negative slope line on the scatter plot (see figure~\ref{fig:codealfa}). Obviously the same effect produces - in the case of source B - a continuum without evident structures below the alpha band.
This feature is potentially dangerous since the purpose of our search is the development of a detector able to identify and reject exactly surface alphas. Indeed, this is the most pernicious background source limiting the sensitivity of the  CUORICINO experiment.
We have observed that this problem appears to be connected to the quality of crystal surface (or to the presence - on the surfaces - of residuals of the abrasive, non-scintillating, powder used for polishing). Indeed the fraction of events populating the tail toward low light emission is different (for the same crystal) if the alpha source hits an optical face of the detector or a rough face.
 Figure~\ref{fig:codealfa} illustrates the test we have done and its results. 
The problem persists, although in reduced size, also in optical crystals as is the case of the ZnSe Huge. 


\subsection{Pulse shape analysis and alpha particle discrimination}\label{sec:PSA}

As discussed in sections~\ref{sec:boloheat} and \ref{sec:bololight} the time response of the bolometers (both the LD and the scintillator crystal), is extremely slow. The light pulse has rise time of the order of few ms that - compared with the typical response of a photomultiplier or a photodiode - is extremely slow. However, at low temperature, also the light emission from scintillators is characterized by long time constants, with different values for alphas or beta/gamma's \cite{tempiscintillazioneabassaT}. This happens also in ZnSe (as evident from figure~\ref{fig:lightpulse}) and a discrimination based on pulse shape analysis is achievable. 

\begin{figure*}
\begin{center}
\includegraphics[width=1\linewidth]{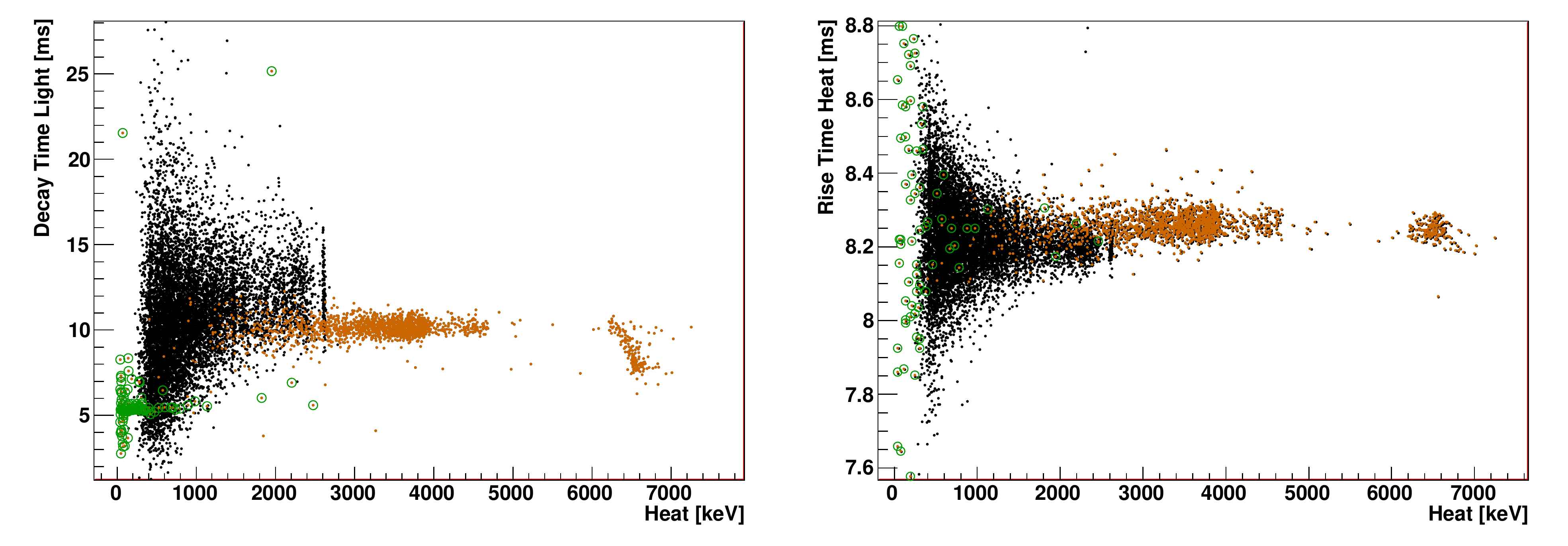}
\end{center}
\caption{Pulse shape of the events represented in the scatter plot of figure~\ref{fig:scatthuge}. Left panel: decay time ($\tau_{decay}$) of light pulses vs. energy ($Heat$) of heat pulses. It is here evident that the three populations identified in figure~\ref{fig:scatthuge} are characterized by different distributions of $\tau_{decay}$. Beta/gamma particle events (black spots) have an average $\tau_{decay}$ of 13 ms, with a wide distribution that partially overlaps with the alpha event distribution. Note that the difference in the width of the alpha and beta/gamma $\tau_{decay}$ distributions is simply due to the different weight of noise on their light pulses. Alphas have a much higher $Light$ amplitude that beta/gamma's, therefore the effect of noise is smaller. This is not evident from the plot since we show $\tau_{decay}$ of light pulse vs. the heat pulse amplitude, ($Heat$). Right panel: rise time ($\tau_{rise}$) of heat pulses vs. energy ($Heat$) for these same events. Alpha and gammas shows different distributions, that however overlap.}
\label{fig:PSA}
\end{figure*}

Figure~\ref{fig:PSA} shows the decay constant (left panel), of events recorded with the LD, versus the energy of the corresponding heat pulse (i.e. $Heat$). This plot refers to the measurement done with the ZnSe Huge crystal, exposed to a \thdt gamma source and the B alpha source, whose scatter plot is shown in figure~\ref{fig:scatthuge}. The three populations identified in figure~\ref{fig:scatthuge} appear to have a clearly different light pulse shape:
\begin{itemize}
\item events of direct ionization on the LD (circles) are very fast, with a decay time of about 5~ms. These events are interaction that involved both the LD and the scintillator. Therefore they are in coincidence with a heat pulse in the scintillator, while the energy recorded by the LD is in part due to direct ionization and in part to scintillation light. Also $^{55}$Fe events have this same decay time (they are not visible in figure~\ref{fig:PSA} because they are not in coincidence with a heat pulse). 
\item alpha events (colored spots) have an average decay time of less than 10 ms and have a distribution quite different from that of beta/gamma events (black spots). Events belonging to the \podd line appear to have a wider distribution of the decay time than events produced by source B alphas. This looks as the same effect that, on the scatter plot of figure~\ref{fig:scatthuge2sources}, makes \podd events extend on a wide range of $Light$ amplitudes. 
\item beta/gamma events (black spots) have an average decay time of less than 13 ms with a wide distribution that is partially superimposed to the alpha one.
\end{itemize}
A similar behaviour is observed on the rise time. \\

Also the shape of heat pulses appears to be different for alpha and beta/gamma events. In the right panel of figure~\ref{fig:PSA} it is shown the rise time of heat pulses. Beta/gammas (black spots) have a faster rise time than alphas (colored spots), the same is true for the decay time.

The reason for having a faster light response and a slower heat one in the case of alphas is not clear and will require dedicated measurements and the construction of a model for signal formation. This will probably help also in the understanding of the other anomalies of this scintillator, discussed so far. 

The observed difference in pulse shapes, open in any case an important possibility for the use of ZnSe bolometers for \BBz search: the rejection of alpha particle background on the basis of the pulse shape.
Indeed, while the use of the $Heat$ to $Light$ ratio seems to provide only a partial discrimination (due to the low $Light$ events discussed in \ref{sec:alphatail}), the use of the pulse shape of light pulses appears much more efficient.

\begin{figure*}
\begin{center}
\includegraphics[width=1\linewidth]{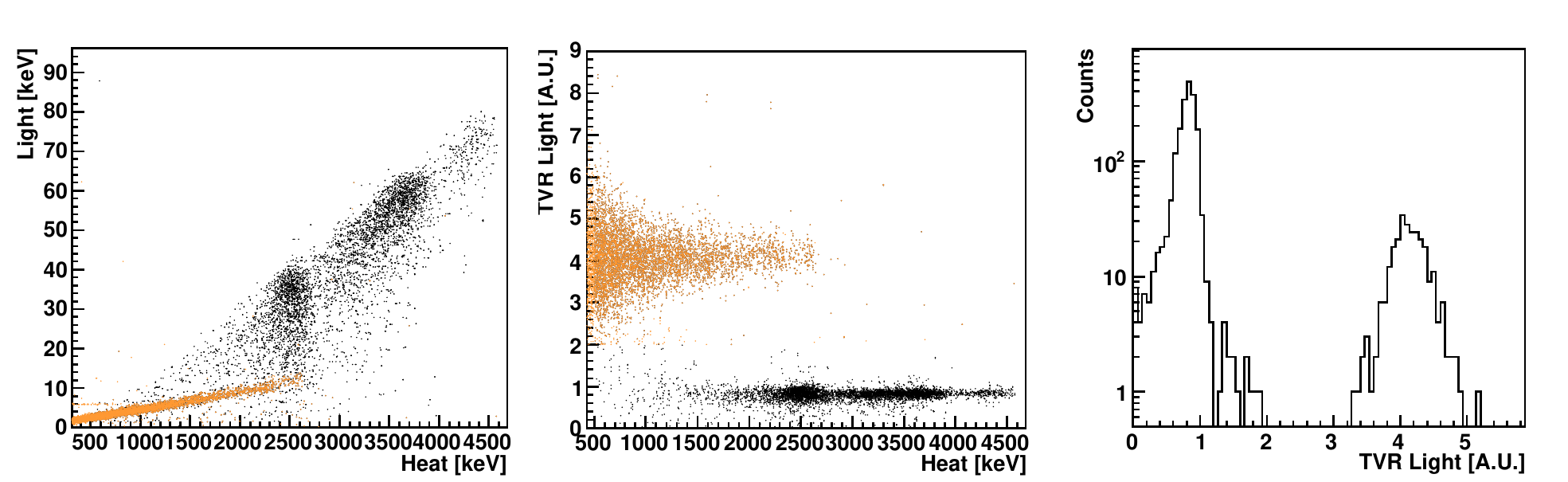}
\end{center}
\caption{ZnSe Large exposed to source B and to a Sm source, both mounted at crystal bottom. Left panel: scatter plot $Light$ vs. $Heat$. Black spots are alpha particles and colored spots beta/gamma's. The distinction is based on a cut defined on the TVR parameter on light pulses. Central panel: TVR distribution for light pulses vs. the amplitude of the corresponding heat pulse ($Heat$). The cut that is used to separate alphas from beta/gamma's is TVR=2. Right panel: TVR distribution for Light pulses between 2400 and 2600~keV.}
\label{fig:discr}
\end{figure*}

In a dedicated run - with optimized measuring conditions - we were able to reach a quite high alpha rejection efficiency.
The ZnSe Large crystal (the one with the best LY) was mounted with - on its bottom - source B and a Sm source ($^{147}$Sm alpha decays with the emission of a 2.2~MeV alpha particle). The LD was operated at a slightly higher temperature than usual, in order to have a faster response. The signal sampling was increased to 16384 points on a 4 sec window, to improve the resolution on the signal shape. 
The result is illustrated in figure~\ref{fig:discr}. The scatter plot (left panel) of $Light$ vs. $Heat$ shows clearly the alpha and beta bands. The alpha band has a quite important tail towards low $Light$ signals and - based on the $Heat$/$Light$ ratio - the alpha discrimination looks quite poor. The LD clearly distinguishes alphas from betas on the basis of their pulse shape (central panel), here measured with the TVR parameter. The projection of the TVR distribution (right panel) shows how efficient can be this kind of discrimination; for a better visualisation of this efficiency all the signals with TVR lower than 2 appear in the scatter plot (left panel) in color.
The alpha rejection efficiency was evaluated in the region 2400-2600 keV (the nearest region to the $^{82}$Se \BB Q-value and - in our measurement - populated enought by both beta/gamma's and alpha's) to be about 15 sigma.

Finally, the possibility of an alpha/beta discrimination based on the scintillating bolometer only (as that presented in the right panel of figure~\ref{fig:PSA}) is extraordinarily interesting since implies an oversimplification for the detector (the LD is no more necessary). However, in our measurements, it is still not enougth sensitive and reproducible. We recall here that ZnSe is not the only scintillating crystal were we have observed this characteristics; a very promising crystal is for example CaMoO$_4$~\cite{gironi}.

\section{\BBz perspectives with ZnSe}\label{sec:dbd}

In this section we briefly discuss the perspectives in the application of ZnSe scintillating bolometers for \BBz study.
 To do this we assume a FWHM energy resolution of 10~keV on the $Heat$ signal at the Se Q-value (\ca2995~keV) and a ROI two FWHM wide (i.e. from 2985 to 3005 keV). For the alpha rejection we assume to use the pulse shape analysis on the light pulse, with the same efficiency reported in section~\ref{sec:PSA} (15 sigma).


In order to translate the results discussed here in the actual sensitivity of a \BB experiment based on this technique, we can refer to the only existing cryogenic setup, the one installed  at LNGS (Hall A) in the 80's by the Milano group and later used also for the CUORICINO experiment. The background model for this setup is in fact well known \cite{ArtRadio} and the extrapolation to the technique described here can be obtained by simply  substituting \teod crystals with ZnSe ones.

For the background rate induced by environmental neutrons and muons we use the results reported in \cite{ArtRadio} and \cite{muCUORICINO}. Indeed the change in the molecular compound does not affect in a relevant way the background rates.  External neutrons yield a contribution integrated in the 3-4 MeV region lower than 10$^{-5}$~c/keV/kg/y \cite{ArtRadio}, while external muons - operating the array in anticoincidence - yield a contribution of about 8~10$^{-5}$~c/keV/kg/y in the 2-4~MeV region \cite{muCUORICINO}. 

For the background induced by environmental gammas, the results reported in \cite{ArtRadio} do not consider the energy region above the 2615~keV line of \tld where the counting rate is drastically reduced (as evident in figure~1 of \cite{ArtRadio}).
In our ROI the dominant contribution comes from \bidq and \tldn. 
Indeed \bidq has few low intensity gamma rays in or above our ROI (the total branching ratio of these gamma lines is 0.08\%) while \tld has no gamma with energy larger than 2615~keV,  but can contribute to the counting rate in our ROI through the contemporary emission of two photons with a sum energy of 3.2~MeV (the 2615~keV and 583~keV gammas).
The background model developed for the Hall A set-up at LNGS, and described in \cite{ArtRadio}, identifies the major source of \bidq and \tld activity as due to contaminations localized in between the external and the internal lead shields. This allows us to evaluate \bidq and \tld contributions in the ROI. These are: \ca 0.006 c/keV/kg/y from \bidq (dominated by the 3000~keV line) and 3~10$^{-5}$ c/keV/kg/y from \tld.

 
Concerning the background contribution of the detector itself, the 15 sigma efficiency in alpha rejection reduce - for reasonable contamination levels of the surfaces and of the crystal bulk - to absolutely negligible values the alpha rate in the ROI. Finally, the use of delayed coincidences allows to get rid of the most worrisome background arising from \udt and \thdt crystal bulk contaminations.
 The measured upper limits on the ZnSe (the Huge crystal) internal contaminations are of 0.4~$\mu$Bq/kg (at 90\% C.L.) for both the two chains; these contaminations results in a background counting rate far below 10$^{-4}$~c/keV/kg/y.  

In conclusion the ultimate sensitivity of a ZnSe based \BBz experiment performed in the Hall A set-up at LNGS would be limited mainly by the \bidq contamination that leads to a background rate in the ROI of \ca0.006 c/keV/kg/y . 

\section{Conclusion}

ZnSe scintillating crystals present a number of anomalies or unexpected behaviours that certainly need further investigation in the future. 
The QF$_{\alpha}$ larger than one could find an explanation in the existence of a light emission from gammas on extremely long time scales (something like an after-glow).
However - even if the usual ratio between LY$_{\alpha}$ and LY$_{\gamma}$ - could be restored with proper collection of all the emitted light, the larger amplitude of alpha particle heat signals with respect to gamma heat signals (see figure~\ref{fig:heatnonlinearity} and section~\ref{sec:boloheat}) remains unexplained. The same holds for the positive slope of gamma monochromatic lines in the scatter plot (and could be also for internal alphas). 
The understanding of these behaviours is obviously mandatory when projecting a \BB experiment since they can influence the response of the detector to a \BBz decay and therefore its efficiency and resolution. 
Nonetheless some conclusion can be drawn concerning the possibility of employing this crystal for a high sensitivity \BBz experiment. The energy resolution - provided that the technique of axis rotation is proven to be reliable also for a \BBz event - is comparable with that of CUORICINO detectors and not very far from Ge diodes. 
As a prove of the perspectives for the application of this detector to future \BBz searches, we have evaluated the background achievable with a medium size experiment in a realistic set-up. We considered the existing cryogenic set-up of Hall A in LNGS that was specially built - in the eighties - for low background experiments. The result is a counting rate of about 6~10$^{-3}$~c/keV/kg/y in the ROI, dominated by the radioactive contamination of the set-up itself. However according to the measured bulk contamination of the ZnSe Huge crystal, the intrinsic background of the detector is lower than 10$^{-4}$ c/keV/kg/y, proving that - in an extremely low background set-up - very high sensitivies for the \BBz process are achievable.

\section {Acknowledgements}
The results reported here have been obtained in the framework of the Bolux R\&D Experiment funded by INFN, aiming at the optimization of a cryogenic DBD technique for a next generation experiment. Thanks are due to E. Tatananni, A. Rotilio, A. Corsi and B. Romualdi  for continuous and constructive help in the  overall setup construction. Finally, we are especially grateful to Maurizio Perego for his invaluable help in the development and improvement  of the Data Acquisition software.


\begin{thebibliography}{00}

\bibitem{reviewDBD}
V. I. Tretyak  and Yu. G. Zdesenko, Atomic Data and Nuclear Data Tables, \rm \bf 80 \rm (2002):83;
\newline S. Elliott and P. Vogel, Ann. Rev. Nucl. Part. Sci. \rm \bf 52 \rm (2002):115;
\newline A. Morales and J. Morales, Nucl. Phys. B (Proc. Suppl.) \rm \bf 114 \rm (2003):141;
\newline F. T. Avignone III, S. R. Elliott and J. Engel (2006), Rev. Mod. Phys. \rm \bf 80 \rm (2008):481;

\bibitem{gerda} 
I. Abt et al. NIM A \rm \bf 570\rm, (2004) 479. S.~Schonert {\it et al.}, Nucl. Phys. (Proc. Suppl.) \rm \bf 145\rm (2005):242. 
\newline I. Abt et al. Eur. Phys. J. C \rm \bf 52\rm (2007):19;

\bibitem{CUORE} C. Arnaboldi {\it et al.}, Astropart. Phys. \rm \bf 20 \rm (2003):91;
\newline C. Arnaboldi {\it et al.}, NIM A \rm \bf 518\rm (2004):775;

\bibitem{majorana} C. E. Aalseth {\it et al.}, Nucl. Phys. B (Proc. Suppl.) \rm \bf 138\rm (2005):217;

\bibitem{exo} M. Danilov {\it et al.}, Phys. Lett. B \rm \bf 480 \rm (2000):12;  

\bibitem{supernemo} I. Nasteva, To be published in the proceedings of EPS-HEP 2009,	arXiv:0909.3167v1;

\bibitem{bolometers}
N.Booth, B. Cabrera and E. Fiorini, Ann. Rev. Nucl. Part. Sci. \rm \bf 46\rm (1996):471;
\newline C. Enss and D. McCammon, J. of Low Temp. Phys.  \rm \bf 151\rm (2008):5;

\bibitem{cuoricino} C. Arnaboldi {\it et al.}, Phys. Rev. Lett. \rm \bf 95\rm (2005):14501;
\newline  C. Arnaboldi {\it et al.}, Phys. Rev. C \rm \bf 78\rm (2008):035502;

\bibitem{NEMOdetector} R. Arnold {\it et al.}, Nucl. Instr. Meth. A \rm \bf 536\rm (2005):79;

\bibitem{nsvecchioarticoloCaF2}
A. Alessandrello {\it et al.}, Nucl. Phys. B (Proc. Suppl.) \rm \bf 28 \rm (1992):233; 
\newline A. Alessandrello {\it et al.}, Phys. Lett. B \rm \bf 420\rm (1998):109;

\bibitem{CRESST} G. Angloher {\it et al.}, Astropart. Phys. \rm \bf 18\rm (2002):1;
\bibitem{ROSEBUD} S. Cebrian {\it et al.}, Phys. Lett. B \rm \bf 563\rm (2003):48;
\bibitem{Pirr06} S.~Pirro {\it et al.}, Phys. Atom. Nucl. \rm \bf 69 \rm (2006):2109;

\bibitem{ArtChambery} M. Pavan , {\it et al.}, Eur. Phys. J A \rm \bf 36\rm (2008):159;
\bibitem{LNGSReport2006} Cuore Collaboration, LNGS Annual Report 2006, www.lngs.infn.it;
\bibitem{paperFrank}C. Arnaboldi {\it et al.}, Phys. Rev. C \rm \bf 78\rm (2008):035502;
\bibitem{Qvalue} G. Audi and A. H. Wapstra, Nucl. Phys. A \rm \bf 729 \rm (2003):337;
\bibitem{NME1} F. Simkovic {\it et al.}, PRC \rm \bf 77  \rm (2008):045503;
\bibitem{NME2} O. Civitarese {\it et al.}, JoP:Conference series \rm \bf 173 \rm (2009):012012;
\bibitem{NME3} J. Menendez {\it et al.}, NPA \rm \bf 818 \rm (2009):139;
\bibitem{NME4} J. Barea and F. Iachello, PRC \rm \bf 79 \rm (2009):044301;
\bibitem{NEMO0n} R. Arnold {\it et al.}, Phys. Rev. Let. \rm \bf 95 \rm (2005):182302 ;
\bibitem{artIoan}I. Dafinei {\it et al.}, IEE Trans. Nucl. Sc. \rm \bf  57 \rm (2010):1;
\bibitem{Landolt-Bornstein} Madelung O., Rassler U., Schulz, M. (ed.). SpringerMaterials - The Landolt-Börnstein Database (http://www.springermaterials.com). DOI: 10.1007/10681719\_480;

\bibitem{luminescenza} E. Krause {\it et al.}, J. Crystal Growth \rm \bf 138 \rm  (1994):75;
\bibitem{scintcresst} H.Krauss, V. B. Mikhailik and D. Wahl, Rad. Meas. \rm \bf 42 \rm (2007):921;
\bibitem{NTD} E. E. Haller {\it et al.}, in “Neutron transmutation doping of semiconducting materials,” edited by R. D. Larrabee (Plenum
Press, New York, 1984), p. 21;
\bibitem{articoloimpulsatore} C. Arnaboldi {\it et al.}, IEEE Trans. Nucl. Sci. \rm \bf 49\rm (2002):2440;
\bibitem{articolostabilizzazione} A. Alessandrello {\it et al.}, Nucl. Instr.  Meth. A \rm \bf 412\rm (1998):454;
\bibitem{elettronica-OF} C. Arnaboldi {\it et al.}, Nucl. Instrum. Meth. A \rm \bf 518 \rm (2004):775;
\bibitem{GattiManfredi} E. Gatti and P.F. Manfredi, Rivista del Nuovo Cimento \rm \bf 9 \rm (1986):1;
\bibitem{paperocdwo4} C. Arnaboldi {\it et al.}, submitted to Astropart. Phys., arXiv:1005.1239v1;
\bibitem{NIMA-2006} S. Pirro, {\it et al.}, Nucl. Instr. and Meth. A \rm \bf 559 \rm (2006):361;
\bibitem{coron} J. Amare {\it et al.}, Applied Physics Letters 87 (2005):264102;
\bibitem{Birks} J.B. Birks, Proc. Phys. Soc. A \rm \bf 64\rm (1951):874;
\bibitem{mather} S.H. Moseley, J.C. Mather, D. McCammon, J. Appl. Phys. \rm \bf 56 \rm 1257:(1984);
\bibitem{Sysoeva} E.V. Sysoeva {\it et al.},  Nucl. Instr. and Meth. A \rm \bf 414 \rm (1997):274;
\bibitem{Klamra} W. Klamra {\it et al.}, Nucl. Instr. and Meth. A \rm \bf 484 \rm (2002):327;
\bibitem{Danevich} F. A. Danevich {\it et al.}, Phys. Rev. C \rm \bf 67 \rm (2003):014310;
\bibitem{tempiscintillazioneabassaT}V. B. Mikhailik, H Kraus, J. Phys. D: Appl. Phys. \rm \bf  39 \rm (2006):1181;
\bibitem{gironi} L. Gironi, Nucl. Instr. Meth. A \rm \bf 617 \rm (2010):478;
\bibitem{ArtRadio} C. Bucci {\it et al.}, Eur. Phys. J. A \rm \bf 41 \rm (2009):155;
\bibitem{muCUORICINO} E. Andreotti {\it et al.}, Astro. Phys., \rm \bf 34 \rm (2010):18.


\end{thebibliography}
\end{document}